\documentclass[oneside,twocolumn,aps,prl,preprintnumbers,
superscriptaddress,amsmath,amssymb]{revtex4-2}

\usepackage{graphicx}
\usepackage{caption}
\usepackage{color}
\usepackage{epstopdf}
\usepackage{amsmath}
\usepackage{braket}
\usepackage{amsthm}
\usepackage{amssymb}
\usepackage{amsfonts}
\usepackage{txfonts}
\usepackage{mathtools}
\usepackage{bm}
\usepackage{hyperref}
\usepackage{lipsum}
\usepackage{natbib}
\usepackage{physics}
\usepackage{cancel}
\usepackage{lipsum}
\usepackage{comment}

\usepackage{subfig}
\usepackage{tikz}
\usetikzlibrary{positioning, arrows.meta, calc}

\usepackage[justification=raggedright,singlelinecheck=false]{caption}

\usepackage{letltxmacro}
\LetLtxMacro{\originaleqref}{\eqref}
\renewcommand{\eqref}{Eq.~\originaleqref}

\newcommand{\Idoperator}{\mathbb{I}}

\newcommand{\re}{{\rm Re}}

\def \ket#1{\mathinner{|{#1}\rangle}}

\newcommand{\hlambda}{\hat{\lambda}}

\newcommand{\hU}{\hat{\mathcal{U}}}

\newcommand{\G}{\mathcal{G}}

\newcommand{\NSIT}{\rm{NSIT}}
\newcommand{\QNSIT}{\rm{QNSIT}}

\newcommand{\Ind}{\rm{Ind}}
\newcommand{\MR}{\rm{MR}}

\newcommand{\bi}{\bar{\imath}}
\newcommand{\bj}{\bar{\jmath}}

\begin{document}

\title{Operational quasiprobability formulation of non-signaling in time for macrorealism violations}

\author{Stefano Gherardini}
\affiliation{Istituto Nazionale di Ottica del Consiglio Nazionale delle Ricerche (CNR-INO), Largo Enrico Fermi 6, I-50125 Firenze, Italy.}
\affiliation{European Laboratory for Non-linear Spectroscopy, Universit\'a di Firenze, I-50019 Sesto Fiorentino, Italy.}

\author{Paolo Solinas}
\affiliation{Dipartimento di Fisica, Universit\`a di Genova, via Dodecaneso 33, I-16146, Genova, Italy.}
\affiliation{INFN - Sezione di Genova, via Dodecaneso 33, I-16146, Genova, Italy.}

\begin{abstract}
In this paper, we introduce the concept of quasi non-signaling in time (QNSIT), which extends non-signaling in time (NSIT) to quasiprobability distributions.
QNSIT requires that the marginals of the quasiprobability distribution under consideration at intermediate times remain unperturbed by the measurement back-action induced by a sequence of quantum measurements. Associated with QNSIT, we propose a quasiprobabilistic characterization of macrorealism, $\MR_{qp}$, which requires that both the QNSIT conditions are satisfied and the quasiprobability distribution constructed from the measurement outcomes is positive. We prove that $\MR_{qp}$ implies NSIT, whereas the converse does not hold because of hidden superposition paths generated or preserved by the system dynamics. We also prove that, for a two-level system, the $\MR_{qp}$ and NSIT conditions coincide. $\MR_{qp}$ therefore provides a more sensitive tool for testing macrorealism and can be implemented experimentally through a quantum non-demolition measurement scheme.
\end{abstract}

\maketitle

{\it Introduction.---}
Under what conditions can a set of measurement records be regarded as compatible with a classical, macrorealist description? To address this question, Leggett and Garg introduced the notion of macrorealism (MR)~\cite{Leggett1985, Emary_2014}. Given a measurement observable with a well-defined spectral decomposition, MR is commonly formulated as the conjunction of three assumptions~\cite{Leggett1985, Emary_2014}: (i) macrorealism \emph{per se} ($\MR_{ps}$), according to which the measured observable has a definite value at all times, thereby excluding coherent superpositions of macroscopically distinct eigenstates~\cite{Halliwell2016, Schmid2024}; (ii) non-invasive measurability (NIM), according to which it is possible, in principle, to determine this value without perturbing the state or its subsequent evolution; and (iii) induction (Ind), according to which future measurements cannot affect the current state of the system~\cite{HalliwellConference2019}.

The physical implications of these assumptions, as well as their experimental implementation, have long been debated~\cite{Knee2012, maroney2014, Knee2016, Halliwell2016, ClementePRL2016}. NIM is arguably the most difficult assumption to verify experimentally: proving NIM requires ruling out the possibility that the observed MR violations arise from measurement-induced disturbances. Because such disturbances cannot be excluded under all experimental conditions, NIM is often replaced by the weaker non-signaling in time (NSIT)~\cite{Kofler2008, Kofler2013, Halliwell2016}. They have a clear operational meaning, since they require the statistical non-invasiveness of the measurement. In other words, even if single measurements can perturb the system's state, the statistical average of the measurement outcomes is not altered by the measurement process, so the marginal statistics remain unperturbed.

Replacing NIM with NSIT conditions offers both practical and theoretical advantages. Experimentally, NSIT can be tested by comparing classical probability distributions built over the measurement outcomes. Theoretically, the violation of an individual Leggett-Garg inequality (LGI) is sufficient to rule out macrorealism, but the validity of the standard LGIs is not generally sufficient to certify MR~\cite{Leggett1985}. By contrast, an appropriate conjunction of NSIT conditions provides a necessary and sufficient condition for the violation of the so-called {\it strong macrorealism} $\MR_{str}$~\cite{Clemente2015, ClementePRL2016, HalliwellConference2019}. Given a set of measurement data obtained from sequential measurements on a system, strong macrorealism establishes whether there exists a classical probability distribution describing the data, whose marginal distributions are unaffected by measurement-induced disturbances. Here, ``classical'' means that the probability distribution satisfies the Kolmogorov axioms of probability.

In this article, we introduce a characterization of MR, $\MR_{qp}$, based on a quasiprobability formulation of the NSIT conditions. Specifically, the validity of $\MR_{qp}$ requires the simultaneous satisfaction of $\MR_{ps}$ and a quasiprobability-based generalization of $\NSIT$, which we refer to as the {\it quasi non-signaling in time} (QNSIT) conditions. Under the assumptions specified below, this criterion is {\it strictly stronger} than the NSIT-based criterion for a macrorealist description of the observed temporal statistics~\cite{Clemente2015, ClementePRL2016, HalliwellConference2019}. In particular, we prove analytically that $\MR_{qp}$ implies $\MR_{str}$. Conversely, by means of a counterexample involving a three-level system, we show that there are situations in which the full set of $\NSIT$ conditions is satisfied, while $\MR_{qp}$ fails because $\MR_{ps}$ is violated. This increased ability to detect violations of MR arises because the $\NSIT$ conditions are blind to superposition paths, developed or preserved by the system's dynamics.

The proposed MR characterization, based on reconstructing a quasiprobability distribution, can be operationally carried out by implementing non-demolition measurements~\cite{SolinasReview2026}. The latter were proposed at the end of 1970s~\cite{ThornePRL1978,Braginsky1980,BraginskyBook,Caves1980,CavesRevModPhys}, and found successful applications in many experimental scenarios, such as atomic magnetometry~\cite{GeremiaPRL2003}, cavity quantum electrodynamics~\cite{GuerlinNature2007}, superconducting digital quantum computing~\cite{solinas2021,solinas2022}, detection of photonic qubits encoded in optical resonators~\cite{NiemietzNature2021} and optical parametric amplifiers for quantum information processing~\cite{YanagimotoPRXQuantum2023}. ND measurements are realized by engineering the coupling between the system and a detector such that the statistics of the measurement observables are preserved, and the $\QNSIT$ conditions are satisfied.

{\it Non-signaling in time conditions.---} 
We briefly detail the definition and meaning of NSIT conditions. Let us consider a system $\mathcal{S}$ that is sequentially measured at times $t_0$, $t_1$ and $t_2$. The result of these measurements is the sets of outcomes $\{a_i\}$ (at $t_0$), $\{b_j\}$ (at $t_1$) and $\{c_k\}$ (at $t_2$), respectively. In between the measurements, the system undergoes some time evolution. As defined in Refs.~\cite{Clemente2015, ClementePRL2016}, we consider a positive joint probability $P(c_k, b_j, a_i)$ to measure the outcomes $a_i, b_j, c_k$. By construction, $P(c_k, b_j, a_i)$ is real, positive and belongs to a {\it classical} probability distribution, such that $\sum_{i,j,k}P(c_k, b_j, a_i)=1$ and Kolmogorov axioms are obeyed. However, the marginal distributions of $P(c_k, b_j, a_i)$ with respect to one or more indices among $i,j,k$ can be affected by the disturbance of the measurement apparatus, in terms of degradation or even cancellation of quantum coherence and correlations. If the measurement apparatus does not {\it statistically} affect the system's state in relation to the observable measured, then NSIT conditions are fulfilled, and the marginal distributions of $P(c_k, b_j, a_i)$ return the probability laws of Born's rule, unperturbed by each intermediate measurement.

Refs.~\cite{Clemente2015,ClementePRL2016} showed that, for three measurement times $t_0,t_1,t_2$, the simultaneous validity of three NSIT conditions provides a necessary and sufficient criterion for the strong macrorealism $\MR_{str}$. Such NSIT conditions are $\NSIT_{(0)12}$, $\NSIT_{0(1)2}$, and $\NSIT_{(1)2}$, where the subscripts indicate the measurement times involved, while parentheses identify the measurement time whose outcome is marginalized over.
These conditions impose constraints on the marginal distributions derived from the joint probability distribution $P(c_k,b_j,a_i)$. Specifically, $\NSIT_{(0)1 2}$ is satisfied if $P(c_k,b_j) = \sum_{i} P(c_k, b_j, a_i)$; $\NSIT_{0(1)2}$ is valid if $P(c_k,a_i) = \sum_{j} P(c_k, b_j, a_i)$; finally, $\NSIT_{(1)2}$ is satisfied if $P(c_k) = \sum_{j} P(b_j, a_i)$, where $P(c_k), P(c_k,a_i), P(c_k,b_j)$ are all unperturbed marginal distributions~\cite{Clemente2015}. 
As discussed above, strong macrorealism entails~\cite{Clemente2015,HalliwellConference2019}:
\begin{equation}\label{eq:NSIT_vs_MRstrong}
\MR_{str} \iff \NSIT_{(0)12} \wedge \NSIT_{0(1)2} \wedge \NSIT_{(1)2} \wedge \Ind.
\end{equation}
As it is customary, throughout this work we assume that the induction condition holds.

At the experimental level, the main difference between the original definition of macrorealism by Leggett and Garg~\cite{Leggett1985, HalliwellConference2019} and the one based on $\NSIT$ conditions is that the former requires certifying {\it a-priori} the non-invasive measurability of the measurement apparatus, while the latter can be tested in any measurement context, even those implementing strong (projective) measurements. For the validity of the $\MR_{str}$, indeed, it is sufficient to verify {\it a-posteriori} that the NSIT conditions are obeyed by analyzing the measurement data $\{ a_i, b_j, c_k \}$. However, notice that, in order to predict which evolutions and measurements fulfill the $\NSIT$ conditions, we need to solve a system of nonlinear equations that increases with the system's size and the number of performed measurements~\cite{HalliwellConference2019, Clemente2015, ClementePRL2016}. Such a system of equations is not guaranteed to admit a solution, and solving it is, in general, computationally expensive.

Furthermore, and more importantly, little emphasis has been placed on the relationship between $\MR_{str}$ and $\MR_{ps}$, namely on the fact that, if $\MR_{str}$ holds, no sequence of evolutions and measurements can give rise to detectable quantum superposition paths. In this regard, the quasiprobability formulation of NSIT conditions introduced below will shed light on this key issue, which is currently of considerable interest for the certification of quantum devices and technologies.

{\it Quasi-non-signaling in time conditions.---} A sequence of measurements, ranging from weak to strong, applied to a quantum system is always invasive, even if only to a minimal extent~\cite{Dicke1981, Peres2000, Busch2009}. This is the reason why the original definition of macrorealism by Leggett and Garg is so challenging to test experimentally.
However, in the spirit of statistical non-invasiveness, there is an alternative route to perform sequential quantum measurements by introducing an auxiliary system that is first coupled to the system and then measured~\cite{GherardiniTutorial, ArvidssonShukur2024review, solinas2024, SolinasReview2026}. Storing information about the system in an auxiliary system can lead to a non-positive-definite statistical distribution, i.e., a quasiprobability distribution, that describes the occurrence of outcomes $\{ a_i, b_j, c_k \}$ at times $t_0, t_1, t_2$. The non-positivity of the quasiprobability distribution can then be used to identify MR violations and, thus, quantum behaviors of the system being measured~\cite{solinas2024, SolinasReview2026}. Notably, these protocols using an auxiliary system do not require any a-posteriori control or solution of a complex system of equations~\cite{Clemente2015, ClementePRL2016, HalliwellConference2019}.

To frame the $\NSIT$ in the quasiprobability language, we introduce the QNSIT conditions. We assume that the data $\{ a_i, b_j, c_k \}$ are, respectively, the eigenvalues of the three system's observables $\hat{A}$ (at time $t_0=0$), $\hat{B}$ (at time $t_1$) and $\hat{C}$ (at time $t_2$), which are sequentially measured. Similar to traditional NSIT~\cite{Kofler2008, Kofler2013, Halliwell2016}, a QNSIT is fulfilled if the marginal distribution of the quasiprobability distribution under scrutiny at some intermediate times is not unperturbed by the back-action due to sequential measurements.

Within the three-measurement setting described above, we introduce the real multi-time quasiprobability density function $\mathcal{P}(\Omega) = \sum_{i,i',j,j',k}q_{k,j,j',i,i'}\,\delta \left(\Omega- f(k,j,j',i,i') \right)$, where $\delta(\cdot)$ is the Dirac delta, $f(k,j,j',i,i') := c_k + (b_j + b_{j'})/2 + (a_i + a_{i'})/2$ and the quasiprobability coefficients $q_{k,j,j',i,i'}$ are defined as
\begin{equation}\label{eq:QP_definition}
    q_{k,j,j',i,i'} := {\rm Tr}\left[ \hat{\Pi}_k^{(C)} \hat{U}_{t_1:t_2} \hat{\Pi}_{j}^{(B)} \hat{U}_{t_0:t_1} \hat{\Pi}_{i}^{(A)} \hat{\rho}_0 \hat{\Pi}_{i'}^{(A)} \hat{U}_{t_0:t_1}^{\dagger} \hat{\Pi}_{j'}^{(B)} \hat{U}_{t_1:t_2}^{\dagger} \right].
\end{equation}
In \eqref{eq:QP_definition}, $\hat{\rho}_0$ is the initial density operator of the system, $\{\hat{\Pi}^{(X)}\}$ is the set of projectors that spectrally decompose the observable $\hat{X}=\hat{A}, \hat{B}, \hat{C}$ at times $t_0,t_1,t_2$, and $\hat{U}_{t_{\ell}:t_{\ell+1}}$ are unitary operators describing the system's dynamics within the time interval $[t_{\ell}:t_{\ell+1}]$ with $\ell\in\{i,j,k\}$. In general, $j'\neq j$ and $i' \neq i$ so that $q_{k,j,j',i,i'}$ is a complex number associated with the time evolution of the system on superposed paths~\cite{solinas2024, SolinasReview2026}; however, the symmetry of (\ref{eq:QP_definition}) guarantees that $\mathcal{P}(\Omega)$ is real.

By construction, the function $f$ is invariant under the independent exchanges $j \leftrightarrow j'$ and $i \leftrightarrow i'$. In addition, for reasons that will become clear below, we further require that $f(k,j,j',i,i') = f(l,r,r',s,s')$ if and only if $l=k$, $\{j,j'\}=\{r,r'\}$, and $\{i,i'\}=\{s,s'\}$, where the braces denote unordered pairs; see the Appendix for details.

In this way, the density function $\mathcal{P}(\Omega)$ fulfills the following properties, which are also discussed in the Appendix:

{\it (i)} Linear decomposition of $\mathcal{P}(\Omega)$ in classical and quantum contributions: 
\begin{equation}
    \mathcal{P}(\Omega) = \mathcal{P}_{\rm cl}(\Omega) + \mathcal{P}_{\rm q}(\Omega).
\end{equation}
The classical contribution $\mathcal{P}_{\rm cl}(\Omega)$ represents the multi-time distribution that is built on the diagonal part $\hat\rho_{t}^{(\rm d)}$ of the system's states at times $t_0,t_1,t_2$, with $\hat\rho_{t} = \hat\rho_{t}^{\rm (d)} + \hat\rho_{t}^{\rm (off-d)}$. $\mathcal{P}_{\rm cl}(\Omega)$ is obtained when the left and right projectors are the same, i.e., $i' = i$ and $j' = j$, whereby $q_{k,j,j,i,i} = P(c_k,b_j,a_i) = P^{(0)}_{i} P^{(1)}_{i \rightarrow j} P^{(2)}_{j \rightarrow k}$. Here, $P^{(0)}_i =  \rho^0_{ii}$, $P^{(1)}_{i \rightarrow j} = |\matrixel{j}{\hat{U}_{t_1:t_2}}{i}|^2$ and $P^{(2)}_{j \rightarrow k} = |\matrixel{k}{\hat{U}_{t_1:t_2}}{j}|^2$, with $P^{(1)}_{i \rightarrow j}, P^{(2)}_{j \rightarrow k}$ transition probabilities. The quantity $q_{k,j,j,i,i}$ is the probability to sequentially record $a_i, b_j, c_k$ at times $t_0, t_1, t_2$ respectively, through a procedure of sequential projective measurements of the observables $\hat{A}, \hat{B}$, and $\hat{C}$~\cite{campisi2011colloquium, Gherardini2022energy}. Hence, $\mathcal{P}_{\rm cl}(\Omega)$ is a positive distribution normalized to one.

Conversely, the quantum contribution $\mathcal{P}_{\rm q}(\Omega)$ is built over the off-diagonal part of $\hat\rho_{t}$, $\hat\rho_{t}^{\rm (off-d)}$, whose elements are $\hat{\Pi}_{\ell}^{(X)}\hat{\rho}_t \hat{\Pi}_{\ell'}^{(X)}$ with $\hat{X}=\hat{A}, \hat{B}, \hat{C}$ and $\ell' \neq \ell$ for any $\ell\in\{i,j\}$.
It turns out that $\mathcal{P}_{\rm q}(\Omega)$ is real but has no definite sign.

{\it (ii)} The integral of $\mathcal{P}_{\rm cl}(\Omega)$ and $\mathcal{P}_{\rm q}(\Omega)$ over $\Omega$ gives $1$ and $0$ respectively:  
\begin{equation}
\int d \Omega ~ \mathcal{P}_{\rm cl}(\Omega)=1 \quad \text{and} \quad \int d \Omega ~ \mathcal{P}_{\rm q}(\Omega)=0.
\end{equation}
Since $\mathcal{P}_{\rm q}(\Omega)$ is real, the latter condition implies that, whenever $\mathcal{P}_{\rm q}(\Omega)$ is not identically zero, it must contain both positive and negative contributions, making $\mathcal{P}(\Omega)$ a quasiprobability distribution density~\cite{SolinasReview2026}.

{\it (iii)} The $\QNSIT$ conditions $\QNSIT_{(0)12}$, $\QNSIT_{0(1)2}$ and $\QNSIT_{(1)2}$ are always satisfied, namely, the marginal distributions are not perturbed by the measurement back-action. Specifically (see also the Appendix), $\QNSIT_{(0)12}$: $q_{k,j,j'} = \sum_{i,i'} q_{k, j,j',i,i'}$; $\QNSIT_{0(1)2}$: $q_{k,i,i'} = \sum_{j,j'} q_{k, j,j',i,i'}$; and $\QNSIT_{(1)2}$: $q_{k} = \sum_{j,j'} q_{k, j,j'}$. This mathematical property is rooted in the requirement that the statistics of the measured observable are not changed by the system-detector coupling. Below, we show that a scheme realizing non-demolition measurements can fulfill this property experimentally.

To formulate a statement similar to \eqref{eq:NSIT_vs_MRstrong} for a quasiprobability density function, we introduce the following quasiprobabilistic MR condition that we denote as $\MR_{qp}$:
\begin{eqnarray}\label{eq:def_MR_qp}
    \MR_{qp} &=& \QNSIT_{(0)12} \wedge \QNSIT_{0(1)2} \wedge \QNSIT_{(1)2}\nonumber \\
    &&\wedge \,\, \mathcal{P}(\Omega) \text{ positive} \wedge \Ind.
\end{eqnarray} 
The macrorealist condition $\MR_{qp}$ is formulated as the traditional definition of macrorealism, i.e., $\MR_{ps} \wedge NIM \wedge \Ind$, except that NIM is substituted with $\QNSIT$. 
As will become clear shortly, the MR condition in Eq.~\eqref{eq:def_MR_qp} was implicitly considered in Refs. ~\cite{solinas2024, SolinasReview2026} in connection with $\MR_{ps}$.

Given the definition in \eqref{eq:QP_definition}, the considered quasiprobability density function $\mathcal{P}(\Omega)$ always satisfies the $\QNSIT$ conditions (i.e., property $(iii)$ above; see also the Appendix), and the MR condition in \eqref{eq:def_MR_qp} simplifies to $\MR_{qp} = \mathcal{P}(\Omega)\text{ positive} \wedge \Ind$. Therefore, as discussed above and in Refs.~\cite{solinas2024, SolinasReview2026}, since $\mathcal{P}(\Omega)\text{ positive} \iff \MR_{ps}$,
\begin{equation}\label{eq:semplified_MR_qp}
    \MR_{qp} = \MR_{ps} \wedge \Ind.
\end{equation}
This means that, by introducing a proper quasiprobability formulation for NSIT conditions, the physical property to be tested for identifying violations of $\MR_{qp}$ is the presence of coherent superpositions, developed or preserved by the system's dynamics, with respect to the measured observables. Such a physical property is tested by looking at the non-positivity of $\mathcal{P}(\Omega)$.

{\it Relation between $\MR_{qp}$ and $\MR_{str}$---} The relation between $\MR_{qp}$ and $\MR_{str}$ of \eqref{eq:NSIT_vs_MRstrong} follows directly from the link existing between the $\NSIT$ and $\QNSIT$ conditions.

\begin{figure}
    \centering
    \includegraphics[width=\columnwidth]{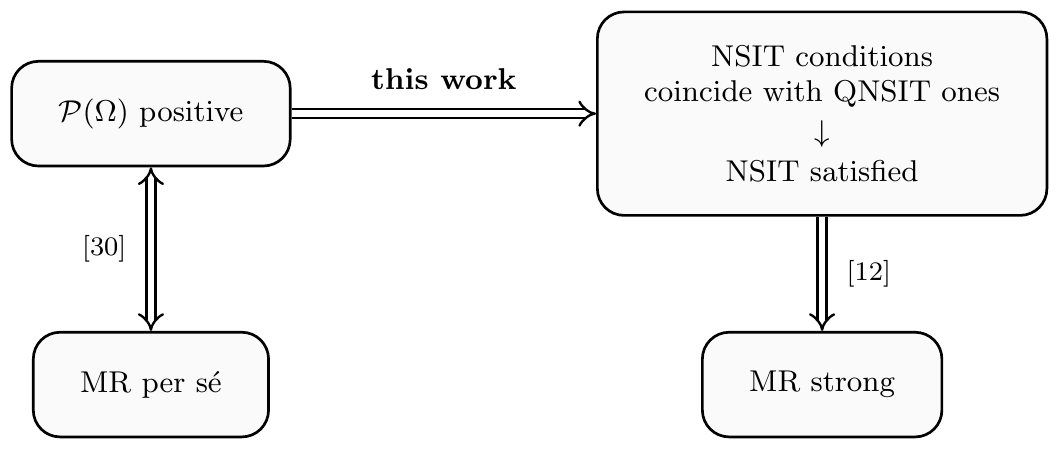}
    \caption{Illustration of the relation between $\MR_{qp}$ and $\MR_{str}$. The positivity of $\mathcal{P}(\Omega)$, achieved if and only if $\MR_{ps}$ is obeyed, entails that the NSIT conditions of \eqref{eq:NSIT_vs_MRstrong} are satisfied, thus certifying the $\MR_{str}$. The reverse is not generally true.}
    \label{fig:scheme}
\end{figure}

Decomposing the $\QNSIT$ conditions in terms of classical and quantum contributions of $\mathcal{P}(\Omega)$, we can prove that if $\mathcal{P}_{\rm q}=0$ (i.e., $\mathcal{P} = \mathcal{P}_{\rm cl}$ and positive), the $\NSIT$ conditions and $\MR_{str}$ are both satisfied.
From the above discussion, this implies that $\MR_{qp} \Rightarrow \MR_{str}$, as illustrated in Fig.~\ref{fig:scheme}.

On the contrary, if the $\NSIT$ conditions are satisfied, then there are cases in which $\mathcal{P}_{\rm q} \neq 0$ that means $\mathcal{P}$ {\it not} positive. Therefore, $\MR_{str} \not \Rightarrow \MR_{qp}$. The corresponding proof and the counterexample with a three-level system, showing the discrepancy between the $\MR_{str}$ and $\MR_{qp}$, are in the Appendix. 

This result shows that, although both $\MR_{str}$ and $\MR_{qp}$ are necessary and sufficient criteria for their respective notions of classicality, $\MR_{qp}$ is more restrictive than $\MR_{str}$. 

There is a clear and physical explanation for such a difference. The $\MR_{str}$ establishes the existence of a classical probability distribution $P(c_k, b_j, a_i)$ that correctly describes the statistics of the sequential measurement outcomes~\cite{Clemente2015, ClementePRL2016, Halliwell2019}. However, since the $\NSIT$ conditions only require the marginals of $P(c_k, b_j, a_i)$ to comply, $\NSIT$ is blind to coherent superpositions of quantum states at times $t_0, t_1, t_2$ that give rise to the same multi-time distribution $P(c_k, b_j, a_i)$. This limitation was already noticed in the original paper~\cite{Clemente2015}.

Instead, the $\MR_{qp}$ certifies the $\MR_{ps}$, in terms of the positivity of $\mathcal{P}(\Omega)$, jointly with the validity of QNSIT conditions at the statistical level. In this regard, the proof in the Appendix clarifies that whenever $\MR_{qp}$ is true and $\mathcal{P}_{\rm q} = 0$, the $\NSIT$ and $\MR_{str}$ are satisfied, and the notion of classicality built on $P(c_k, b_j, a_i)$ is respected. On the contrary, there are processes that develop and/or leave unaltered quantum coherences/correlations (thus, violating the $\MR_{ps}$) that {\it statistically} appear as if obtained from a classical distribution.

Finally, as an interesting observation, in the Appendix we show that for a two-level system the two macrorealistic conditions coincide: $\MR_{qp} {\iff} \MR_{str}$. This follows from the impossibility, for a qubit-like system, of supporting dark subspaces in which quantum coherences are protected from the back-action of the measurement apparatus. These dark subspaces refer to the system's Hilbert space projected onto the eigenspaces of the measured observables.

{\it Non-demolition measurements for operational QNSIT.---}The non-demolition (ND) quasiprobability distribution discussed in Refs.~\cite{solinas2015fulldistribution,solinas2021,solinas2022} respects all properties {\it (i)-(iii)}, and thus represents the ideal tool for experimental applications. ND measurements are sequential non-projective measurements, and are realized by engineering the coupling between the system and a detector at times $t_0, t_1, t_2$ such that the $\QNSIT$ conditions are satisfied. Eventually, the accumulated phase in the detector's state is measured by an interferometric method that is able to extract information about the system. Details about the ND-measurement scheme are discussed in Refs.~\cite{solinas2024, melegari2025, SolinasReview2026}; thus, we only recall its essential properties.

Operationally, consider the same system $\mathcal{S}$ as above, and a detector $\mathcal{D}$. We denote by $\hlambda$ the operator acting on the degrees of freedom of $\mathcal{D}$. The system-detector interaction is governed by the unitary transformation $\hU_{g} = \exp{ig(\hat{X} \otimes \hlambda)}$. This choice ensures that the statistics of $\hat{X}=\hat{A}, \hat{B}, \hat{C}$ is not modified by the coupling, such that we are effectively performing a statistical non-invasive measurement and storing the desired information in the detector's phase. The initial state of the detector is taken as the superposition $\ket{\phi} = (\ket{\lambda} + \ket{-\lambda})/\sqrt{2}$ of two $\hlambda$'s eigenstates, with $\hlambda \ket{\pm \lambda} = \pm \lambda \ket{\pm\lambda}$.

By measuring the accumulated phase of the detector's state as a function of the coupling strength $g$, we obtain the characteristic function $\G_g$, whose Fourier transform yields the quasiprobability density function $\mathcal{P}(\Omega)$ constructed from the coefficients $q_{k,j,j',i,i'}$ defined in \eqref{eq:QP_definition}. In principle, any scheme that realizes quantum ND measurements and is associated with the quasiprobability coefficients in \eqref{eq:QP_definition}, is an appropriate method to certify violations of MR, provided it allows for identifying the presence of quantum superposition paths along the basis defined by the measurement observables.

{\it Conclusions.---} In this article, we have extended the notion of non-signaling in time to quasiprobability distributions, thereby proposing a novel MR condition that we denote by $\MR_{qp}$. We have proved that $\MR_{qp}$ is stricter than $\MR_{str}$, as it can also detect the presence of superposition paths---developed or preserved by the system's dynamics---that nevertheless satisfy the $\NSIT$ conditions. We also show that, for two-level systems, the $\MR_{qp}$ and $\MR_{str}$ coincide.

Operationally (e.g., implementing non-demolition measurements), observing negative regions in $\mathcal{P}(\Omega)$ signals the presence of quantum behaviors by the system. In fact, if $\mathcal{P}(\Omega)$ is positive, then both $\MR_{qp}$ and $\MR_{str}$ are satisfied, and the system behaves classically. Conversely, if $\mathcal{P}(\Omega)$ has negative regions, then the $\MR_{qp}$ is violated, and a check of the $\MR_{str}$ identifies the source of the violation. When the $\MR_{str}$ is satisfied, the violation of MR depends on underlying quantum dynamics allowing for superposition paths, while if $\MR_{str}$ is not fulfilled, then also the $\NSIT$ conditions are violated.

As an outlook, in order to make the MR formulation with quasiprobabilities easily testable with an experimental platform severely exposed to external noise, it is desirable to extend the demonstrations presented here to generic completely positive, trace-preserving open quantum maps, still under the assumption of operating with sequential measurements. Furthermore, it is worth studying how the quasiprobabilistic MR condition $\MR_{qp}$ [\eqref{eq:def_MR_qp}] relates to the necessary and sufficient criteria based on a suitable augmented set of LGIs~\cite{Halliwell2017, HalliwellConference2019}. Finally, one might also investigate how the quasiprobability coefficients $q_{k,j,j',i,i'}$ of \eqref{eq:QP_definition} have to be modified to include sequential weak measurements~\cite{MitchisonPRA2007, PiacentiniPRL2016, ThekkadathPRL2016, AvellaPRA2017} or extended Kirkwood-Dirac quasiprobability approaches taking into account temporal quantum states~\cite{LiePRL2025, JiaArxiv2026}.

\begin{acknowledgments}
{\it Acknowledgements.---} S.G.~acknowledges financial support from the PRIN project 2022FEXLYB Quantum Reservoir Computing (QuReCo), and the PNRR MUR project PE0000023-NQSTI funded by the European Union---Next Generation EU. 

{\it Authors' contribution---} The authors contributed equally to this work.
\end{acknowledgments}

\appendix*
\section{APPENDIX}

\section{Constraints on the function $f$} 

In the main text, we have assumed that $f(k,j,j',i,i') = c_k + (b_j+b_{j'})/2 + (a_i+a_{i'})/2$ satisfies the condition $f(k,j,j',i,i') = f(l,r,r',s,s')$ only if $l=k$, $\{j,j'\} = \{r,r'\}$ and $\{i,i'\} = \{s,s'\}$, where braces denote unordered pairs, including multiplicities. By construction, $f$ is invariant under the independent exchanges $j \leftrightarrow j'$ and $i \leftrightarrow i'$. The assumption excludes any additional coincidences between recorded values of $f$, i.e., $f(k,j,j',i,i') \ne f(l,i,i',j,j')$ for $k \neq l$ or $\{j,j'\} \neq \{i,i'\}$. Trajectories with different final indices $k$ or different unordered index pairs $\{j,j'\}$ and $\{i,i'\}$ are related with Dirac delta functions with different arguments. This separation is the key to prove that vanishing $\mathcal P_{\rm q}$ implies that $\NSIT$ conditions are valid.

As a practical example, consider the scheme realizing quantum non-demolition measurements~\cite{solinas2024, SolinasReview2026}. In addition to the system-detector coupling $\exp\left\{ ig \hat{A} \otimes \hat{\lambda} \right\}$ with $\hat{A}$ and $\hat{\lambda}$ acting on the system's and detector's degrees of freedom, we apply $\exp\left\{ ig \, \alpha_{i}\hat{\Pi}^{(A)}_i \otimes \hat{\lambda} \right\}$ at the same measurement time, with tunable real parameters $\alpha_i$. Here, $\hat{\Pi}^{(A)}_i = |a_i\rangle\!\langle a_i|$ is the projector onto the eigenstate $\ket{a_i}$ of $\hat{A}$. Since these couplings commute, their combined action is $\exp \Big\{ ig \widetilde{\hat{A}} \otimes \hat{\lambda} \Big\}$, where $\widetilde{\hat{A}} = \sum_i (a_i+\alpha_i)\hat{\Pi}^{(A)}_i$. Hence, the accumulated phase on the state $\ket{a_i}\otimes\ket{\lambda}$ is $\exp\Big\{ ig \lambda (a_i+\alpha_i) \Big\}$.
Analogous couplings at the other measurement times replace $b_j$ by $b_j+\beta_j$ and $c_k$ by $c_k+\gamma_k$. Choosing $\alpha_i$, $\beta_j$ and $\gamma_k$ avoiding those constraints ensures the separation condition stated above.

\section{{\rm NSIT} and {\rm QNSIT} conditions}

For the sake of the discussion, we use a slightly simplified notation for the indices. We denote the classical probabilities as $P(c_k, b_j, a_i) \equiv P(k, j, i)$, and we consider that
$q_{k, j,j',i,i'} \equiv {\rm Tr}\left[ \hat{\Pi}_k \hat{U}_{t_1:t_2} \hat{\Pi}_{j} \hat{U}_{t_0:t_1} \hat{\Pi}_{i} \hat{\rho}_0 \hat{\Pi}_{i'} \hat{U}_{t_0:t_1}^{\dagger} \hat{\Pi}_{j'} \hat{U}_{t_1:t_2}^{\dagger} \right]$, where we have suppressed the observable labels on the spectral projectors defining $\hat{\Pi}_k^{(C)}\equiv\hat{\Pi}_k$, $\hat{\Pi}_j^{(B)}\equiv\hat{\Pi}_j$ and $\hat{\Pi}_i^{(A)}\equiv\hat{\Pi}_i$.

With the assumptions on $f(k, j,j',i,i')$ and the symmetry $q^{*}_{k, j',j,i',i} = q_{k, j,j',i,i'}$, the quantum contributions to $\mathcal{P}(\Omega)$ is~\cite{SolinasReview2026}
\begin{eqnarray*}\label{eq:P_q_full}
&&\mathcal{P}_{\rm q}(\Omega)
= 2\sum_{k,j}\sum_{i<i'}
\re[q_{k,j,j,i,i'}] \, \delta[\Omega-f(k,j,j,i,i')] \nonumber\\
&& + 2\sum_{k,i}\sum_{j<j'}
\re[q_{k,j,j',i,i}] \, \delta[\Omega-f(k,j,j',i,i)] \nonumber\\
&& + 2\sum_k\sum_{\substack{i<i' \\ j<j'}}
\re[q_{k,j,j',i,i'}+q_{k,j,j',i',i}]
\, \delta[\Omega-f(k,j,j',i,i')].
\end{eqnarray*}
Since $\mathcal P_{\rm q}$ is real and satisfies $\int d\Omega\,\mathcal P_{\rm q}(\Omega)=0$, any nonzero quantum contribution necessarily produces a negative weight for a point in the support of the full distribution~\cite{solinas2024,SolinasReview2026}. Consequently, $\mathcal P$ is non-negative if and only if $\mathcal P_{\rm q} = 0$, where
\begin{equation}\label{eq:P_q_zero}
\mathcal{P}_{\rm q}\equiv0 \iff
\begin{cases}
\re[q_{k,j,j',i,i}]=0
& \forall\,k,i,\ j<j',\\
\re[q_{k,j,j,i,i'}]=0
& \forall\,k,j,\ i<i',\\
\re[q_{k,j,j',i,i'}+q_{k,j,j',i',i}]=0
& \forall\,k,\ i<i',\ j<j'.
\end{cases}
\end{equation}

The classical probability for three sequential measurements is obtained when $i'=i$ and $j'=j$, i.e., $P(k, j,i) \equiv q_{k, j, j, i, i}$ with
\begin{equation*}
    P(k, j,i) = {\rm Tr}\left[ \hat{\Pi}_k \hat{U}_{t_1:t_2} \hat{\Pi}_{j} \hat{U}_{t_0:t_1} \hat{\Pi}_{i} \hat{\rho}_0 \hat{\Pi}_{i} \hat{U}_{t_0:t_1}^{\dagger} \hat{\Pi}_{j} \hat{U}_{t_1:t_2}^{\dagger} \right].
\end{equation*}
Following Ref.~\cite{Halliwell2016}, the two-time probability $P(j,i)$ for measurements at times $t_0$ and $t_1$ is 
\begin{eqnarray*}
    P(j,i) = {\rm Tr}\left[   \hat{\Pi}_{j} \hat{U}_{t_0:t_1} \hat{\Pi}_{i} \hat{\rho}_0 \hat{\Pi}_{i} \hat{U}_{t_0:t_1}^{\dagger} \right].
\end{eqnarray*}
In an analogous way, we can write
\begin{eqnarray*}
    P(k,j) &=& {\rm Tr}\left[ \hat{\Pi}_k \hat{U}_{t_1:t_2} \hat{\Pi}_{j} \hat{U}_{t_0:t_1}  \hat{\rho}_0  \hat{U}_{t_0:t_1}^{\dagger} \hat{\Pi}_{j} \hat{U}_{t_1:t_2}^{\dagger} \right] \nonumber \\
    P(k,i)
    &=& {\rm Tr}\left[\hat{\Pi}_k\hat{U}_{t_1:t_2}\hat{U}_{t_0:t_1}
    \hat{\Pi}_i\hat{\rho}_0\hat{\Pi}_i \hat{U}_{t_0:t_1}^{\dagger}\hat{U}_{t_1:t_2}^{\dagger} \right]  \\
    P(k) &=& {\rm Tr}\left[ \hat{\Pi}_k \hat{U}_{t_1:t_2}  \hat{U}_{t_0:t_1}  \hat{\rho}_0  \hat{U}_{t_0:t_1}^{\dagger}  \hat{U}_{t_1:t_2}^{\dagger} \right],\nonumber
\end{eqnarray*}
which correspond, respectively, to measurements only at times $t_1$ and $t_2$, only at times $t_0$ and $t_2$, and at time $t_2$. With these definitions, the $\NSIT$ conditions read as~\cite{Clemente2015, ClementePRL2016, Halliwell2016} 
\begin{eqnarray}\label{eq:NSIT_conditions}
     \NSIT_{(1)2} = {\rm true} &\iff&  P(k) = \sum_{j} P(k, j) \nonumber \\
     \NSIT_{(0)12} = {\rm true} &\iff&  P(k,j) = \sum_i P(k, j,i) \\
     \NSIT_{0(1)2} = {\rm true} &\iff&  P(k,i) = \sum_j P(k, j,i). \nonumber
\end{eqnarray}

The $\QNSIT$ conditions are defined following a similar reasoning, but taking into account also the cases $i'\neq i$ and $j'\neq j$.
For example, if we marginalize $q_{k, j,j',i,i'}$ over the initial measurement at time $t_0$, the corresponding reduced coefficient is $q_{k,j, j'} = {\rm Tr}\left[  \hat{\Pi}_k \hat{U}_{t_1:t_2} \hat{\Pi}_{j} \hat{U}_{t_0:t_1}  \hat{\rho}_0 \hat{U}_{t_0:t_1}^{\dagger}  \hat{\Pi}_{j'} \hat{U}_{t_1:t_2}^{\dagger} \right]$. Hence, the condition $\QNSIT_{(0)12}$ becomes:
\begin{eqnarray} \label{eq:QNSIT_{(0)12}}
    \QNSIT_{(0)12} = {\rm true} &\iff&  q_{k, j,j'} = \sum_{i,i'} q_{k, j,j',i,i'}.
\end{eqnarray}
Similarly we denote $q_{k,i,i'}$ as the reduced coefficient obtained by marginalizing $q_{k, j,j',i,i'}$ over the measurement at $t_1$, and we set $q_k=P(k)$. We thus find:
\begin{eqnarray} \label{eq:QNSITs}
     \QNSIT_{0(1)2} = {\rm true} &\iff&  q_{k, i,i'} = \sum_{j,j'} q_{k, j,j',i,i'} \nonumber \\
     \QNSIT_{(1)2}  = {\rm true} &\iff&  q_{k} = \sum_{j,j'} q_{k, j,j'}.
\end{eqnarray}

By using the completeness and orthogonality relations of the projectors, i.e., $\sum_p \hat{\Pi}_p^{(X)} = \hat{\Idoperator}$, $\left(\hat{\Pi}_p^{(X)}\right)^2 = \hat{\Pi}_p^{(X)}$ and $\hat{\Pi}_p^{(X)} \hat{\Pi}_m^{(X)} = \delta_{pm} \hat{\Pi}_m^{(X)}$ (for $X=A, B, C$), it can be checked that all $\QNSIT$ conditions are satisfied by the quasiprobability $q_{k, j,j',i,i'}$ (see also Refs~\cite{solinas2024, SolinasReview2026}). 
For example, $\sum_{i,i'} q_{k,j,j',i,i'} =  {\rm Tr}\left[ \hat{\Pi}_k \hat{U}_{t_1:t_2} \hat{\Pi}_{j} \hat{U}_{t_0:t_1}  \hat{\rho}_0 \hat{U}_{t_0:t_1}^{\dagger} \hat{\Pi}_{j'} \hat{U}_{t_1:t_2}^{\dagger} \right] = q_{k, j, j'}$, meaning that the $\QNSIT_{(0)12}$ is satisfied. Analogous calculations can be done for $\QNSIT_{0(1)2}$ and $\QNSIT_{(1)2}$, thus verifying that QNSIT relations are always satisfied by $q_{k, j, j', i, i'}$.

\section{Relation between NSIT and QNSIT}

QNSIT relations (\ref{eq:QNSIT_{(0)12}})-(\ref{eq:QNSITs}) contain quantum contributions that are not present in the usual NSIT conditions (\ref{eq:NSIT_conditions}).
To highlight the relation between them, we isolate the quantum contributions.

For the $\QNSIT_{(1)2}$, the classical contributions are given by $i'=i$ and $j'=j$, while the quantum contributions correspond to $i'\neq i$ or $j' \neq j$. Using the symmetry of the quasiprobabilities $q_{k, j,j',i,i'}$, we can write that 
\begin{eqnarray*}
q_k &=& P(k)
= \sum_{i,j} q_{k,j,j,i,i} + 2\sum_j\sum_{i<i'} \re[q_{k,j,j,i,i'}] \nonumber\\
&+& 2\sum_i\sum_{j<j'} \re[q_{k,j,j',i,i}] 
+ 2\sum_{i<i',j<j'}
\re[q_{k,j,j',i,i'}+q_{k,j,j',i',i}].
\end{eqnarray*}
Hence, as $q_k=P(k)$ and $q_{k,j,j,i,i}=P(k,j,i)$, one has:
\begin{eqnarray}\label{eq:(1)2_relation}
&& P(k) - \sum_{i,j} P(k,j,i) = 2\sum_j\sum_{i<i'} \re[q_{k,j,j,i,i'}] \nonumber \\
&& + 2\sum_i\sum_{j<j'} \re[q_{k,j,j',i,i}] + 2\sum_{i<i',j<j'}
\re[q_{k,j,j',i,i'}+q_{k,j,j',i',i}].\nonumber \\
\end{eqnarray}

For the $\QNSIT_{(0)12}$ in \eqref{eq:QNSITs}, which we aim to connect with the $\NSIT_{(0)12}$, we consider $j'=j$ and we separate the classical contributions given by $i'=i$ from the quantum ones with $i'\neq i$. We thus have: $q_{k,j,j} = P(k,j) = \sum_{i} q_{k,j,j,i,i} + 2 \sum_{i<i'} \re[q_{k, j,j , i,i'}]$, where $\sum_{i} q_{k,j,j,i,i} = \sum_i P(k,j,i)$ represents the classical contribution. As a result,  
\begin{eqnarray} \label{eq:(0)12_relation}
    P(k,j) - \sum_i P(k,j,i) =  2 \sum_{i<i'} \re[q_{k, j,j , i,i'}].
\end{eqnarray}

Finally, for the $\QNSIT_{0(1)2}$, we consider $i'=i$ and $q_{k,i,i} = P(k,i) = \sum_{j} q_{k,j,j,i,i} + 2 \sum_{j< j'} \re[q_{k, j, j' , i,i}]$, i.e.,
\begin{eqnarray} \label{eq:0(1)2_relation}
    P(k,i) - \sum_j P(k,j,i) =  2 \sum_{j< j'} \re[q_{k, j, j' , i,i}].
\end{eqnarray}

The right-hand sides of Eqs.~(\ref{eq:(1)2_relation}), (\ref{eq:(0)12_relation}), (\ref{eq:0(1)2_relation}) are directly connected to the $\NSIT$ conditions in \eqref{eq:NSIT_conditions}. In particular, under the separation assumption on $f$, $\mathcal P_{\rm q}=0$ forces all the grouped coefficients in \eqref{eq:P_q_zero} to vanish. Then, Eqs.~(\ref{eq:(0)12_relation})-(\ref{eq:0(1)2_relation}) imply $\NSIT_{(0)12}$ and $\NSIT_{0(1)2}$, respectively. Moreover, if the right-hand side of
\eqref{eq:(1)2_relation} is vanishing, then $P(k) = \sum_{i,j}P(k,j,i)$.
Combining this equality with the already established $\NSIT_{(0)12}$ yields $P(k)=\sum_j\left[\sum_i P(k,j,i)\right] =\sum_j P(k,j)$, which entails $\NSIT_{(1)2}$. Hence, all the three NSIT conditions hold, as well as the $\MR_{str}$. Finally, recalling $\mathcal{P}_{\rm q} = 0 \iff \MR_{qp}$, we obtain the implication
\begin{equation*}\label{eq:M_qp_implies_MR_str}
    \MR_{qp} \Rightarrow \MR_{str}.
\end{equation*}

The converse is not generally true in dimensions higher than two. In fact, the validity of $\NSIT$ conditions only constrains sums of the real part of $q_{k, j,j',i,i'}$; whereas, for $\mathcal P_{\rm q}$ to be vanishing, each term in \eqref{eq:P_q_zero} must be equal to zero separately. For example, $\NSIT_{0(1)2}$ requires $\sum_{j<j'}\re[q_{k,j,j',i,i}]=0$ for every $k,i$. However, in $\mathcal P_{\rm q}$, the quasiprobabilities in such a condition multiply different Dirac delta functions. Hence, their sum can vanish without their contribution to $\mathcal P_{\rm q}$ is vanishing.

In order to further clarify this point, we discuss an example where all three NSIT conditions are simultaneously valid, while $\mathcal P_{\rm q} \ne 0$.
Consider a three-level system with measurement projectors onto the computational basis $\{\ket{0},\ket{1},\ket{2}\}$ at all three times. We choose the encoded spectra $(a_0,a_1,a_2)=(0,1,3)$, $b_j=7a_j$ and $c_k=49a_k$, which satisfy the separation condition on $f$. The initial density matrix of the system is 
\begin{equation*}
    \hat\rho_0 = 
    \left(
\begin{array}{ccc}
 p_0 & \epsilon  & -\epsilon  \\
 \epsilon  & p_1 & 0 \\
 -\epsilon  & 0 & p_2 \\
\end{array}
\right),
\end{equation*}
with $p_i > 0$, $\sum_i p_i =1$, $\epsilon\in\mathbb R\,:\,0<|\epsilon|\leq\sqrt{p_0p_1p_2/(p_1+p_2)}$. We choose $\theta>0$, $\phi < \frac{\pi}{2}$ in order to have nonzero terms in $\mathcal{P}_{\rm q}$.

Denoting with $\hat{\sigma}^k_{ij}$ the k-Pauli matrix acting on the sectors $\ket{i}$ and $\ket{j}$ of the system's Hilbert space, we take $\hat{U}_{t_0:t_1} = e^{i \theta \hat{\sigma}^y_{01}} e^{i \frac{\pi}{4} \hat{\sigma}^y_{12}}$ and $\hat{U}_{t_1:t_2} = e^{i \phi \hat{\sigma}^x_{01}}$. Under these conditions, it can be verified that all the $\NSIT$ conditions above are satisfied.
However, by direct calculation, it turns out that 
\begin{eqnarray*}
    q_{0, 0, 0, 1,0} + q_{0, 0, 0, 0,1} &=& \sqrt{2} \epsilon  \sin(\theta) \cos(\theta)\cos^2(\phi) \nonumber \\
    q_{0, 0, 0, 2,0} + q_{0, 0, 0, 0,2} &=& -\sqrt{2} \epsilon  \sin(\theta)\cos(\theta)\cos^2(\phi).
\end{eqnarray*}
Although $\sum_{i\ne i'}q_{0,0,0,i,i'}=0$, the two separated grouped quantities are nonzero and multiply $\delta(\Omega-1/2)$ and $\delta(\Omega-3/2)$, respectively. The separation condition on $f$ excludes the cancellation with other groups. Consequently, $\mathcal P_{\rm q}\ne 0$, even though all three NSIT conditions hold. As a consequence, $\mathcal{P}_{\rm q} \neq 0$, and $\MR_{qp} = \MR_{ps}$ is violated.

\section{$\MR_{qp}{\iff}\MR_{str}$ for two-level systems}

For a two-level system, each index $i$ or $j$ has a unique alternative value: $\bi = 1-i$ and $\bj = 1-j$, respectively. By exploiting the invariance of $f$ under exchanges $i\leftrightarrow i'$ and $j\leftrightarrow j'$, $\mathcal{P}_{\rm q}(\Omega)$ is
\begin{eqnarray}\label{eq:P_q_two-level}
&&\mathcal{P}_{\rm q}(\Omega)
= 2 \sum_{k,i}
\re [q_{k,0,1,i,i}]
\, \delta[\Omega-f(k,0,1,i,i)] \nonumber\\
&&+ 2 \sum_{k,j}\re [q_{k,j,j,0,1}]
\, \delta[\Omega-f(k,j,j,0,1)] \nonumber \\
&&+ 2 \sum_k \left( \re [q_{k,0,1,0,1}]  + \re [q_{k,0,1,1,0}]\right)
\, \delta[\Omega-f(k,0,1,0,1)]. \nonumber \\
\end{eqnarray}
From \eqref{eq:0(1)2_relation}, the validity of $\NSIT_{0(1)2}$ implies $\sum_j \re[q_{k,0,1,i,i}]=0$ for every $k,i$. Hence, the first contribution of \eqref{eq:P_q_two-level} vanishes. Similarly, from \eqref{eq:(0)12_relation}, if $\NSIT_{(0)12}$ is satisfied, then $\sum_i \re[q_{k,j,j,0,1}]=0$ for every $k,j$, so that also the second contribution of \eqref{eq:P_q_two-level} goes to zero. Furthermore, the joint validity of $\NSIT_{(1)2}$ and $\NSIT_{(0)12}$ gives $P(k)=\sum_{i,j}P(k,j,i)$. Hence, since the first two terms in the right-hand-side of \eqref{eq:(1)2_relation} already vanish, \eqref{eq:(1)2_relation} implies that $\re [q_{k,0,1,0,1}]  + \re [q_{k,0,1,1,0}]=0$ for every $k$. As a result, for a two-level system, if the three $\NSIT$ conditions are all satisfied, then $\mathcal P_{\rm q}=0$, which establishes $\MR_{str}\Rightarrow\MR_{qp}$. Combining this result with the reverse implication established before, we obtain $\MR_{qp} {\iff} \MR_{str}$. 

\bibliographystyle{apsrev}
\bibliography{QuasiProbabilities}

\end{document}